\documentclass[showpacs,twocolumn,floatfix,amsmath,amssymb,superscriptaddress,prb,nopacs]{revtex4-2}
\usepackage{graphicx}
\usepackage{float}
\usepackage{dcolumn}
\usepackage{bm}
\usepackage{amssymb}
\usepackage{amsmath}
\usepackage{hyperref} 
\hypersetup{bookmarks=true,unicode=true,colorlinks=true, urlcolor=blue,linkcolor=blue,citecolor=blue,filecolor=blue}

\newcommand{\bt}{\textbf}
\usepackage{physics}
\usepackage{comment}
\graphicspath{{figures/}}
\usepackage{float}
\usepackage{subfigure}
\DeclareMathOperator{\arcsinh}{arcsinh}

\begin{document}

\title{Ballistic transport through quantum point contacts of multi-orbital oxides}

\author{J. Settino}
\affiliation{CNR-SPIN c/o Universit\'a degli Studi di Salerno, I-84084 Fisciano (Sa), Italy}
%

%
\author{C. A. Perroni}
\affiliation{CNR-SPIN c/o
Universit\'a degli Studi di Napoli Federico II,
\\ Complesso Universitario Monte S. Angelo, Via Cintia, I-80126 Napoli, Italy}
\affiliation{Physics Department "Ettore Pancini",
Universit\'a degli Studi di Napoli Federico II,
\\ Complesso Universitario Monte S. Angelo, Via Cintia, I-80126 Napoli, Italy}

\author{F. Romeo}
\affiliation{Dipartimento di Fisica ``E. R. Caianiello", Universit\'a degli Studi di Salerno, I-84084 Fisciano (Sa), Italy}
\author{V. Cataudella}
\affiliation{CNR-SPIN c/o
Universit\'a degli Studi di Napoli Federico II,
\\ Complesso Universitario Monte S. Angelo, Via Cintia, I-80126 Napoli, Italy}
\affiliation{Physics Department "Ettore Pancini",
Universit\'a degli Studi di Napoli Federico II,
\\ Complesso Universitario Monte S. Angelo, Via Cintia, I-80126 Napoli, Italy}
\author{R. Citro}
\affiliation{CNR-SPIN c/o Universit\'a degli Studi di Salerno, I-84084 Fisciano (Sa), Italy}
\affiliation{Dipartimento di Fisica ``E. R. Caianiello", Universit\'a degli Studi di Salerno, I-84084 Fisciano (Sa), Italy}
\affiliation{INFN, Sezione di Napoli, 80126 Napoli NA, Italy}

\begin{abstract}
Linear and non-linear transport properties through an atomic-size point contact based on oxides two-dimensional electron gas is examined using the tight-binding method and the $\mathbf{k\cdot p}$ approach. The ballistic transport is analyzed in contacts realized at the (001) interface between  band insulators $LaAlO_3$ and $SrTiO_3$ by using the Landauer-B\"uttiker method for many sub-bands derived from three Ti 3d orbitals ($d_{yz}$, $d_{zx}$ and $d_{xy}$) in the presence of an out-of-plane magnetic field.  
We focus especially on the role played by the atomic spin-orbit coupling and the inversion symmetry breaking term pointing out three transport regimes: the first, at low energies, involving the first $d_{xy}$-like sub-bands, where the conductance quantization is robust; a second one, at intermediate energies, entailing further $d_{xy}$-like sub-bands, where the sub-band splitting induced by the magnetic field is quenched; the third one, where the mixing between light $d_{xy}$-like, heavy $d_{yz}$-like and $d_{zx}$-like sub-bands is so strong that the conductance plateaus turn out to be very narrow. Very good agreement is found with recent experiments exploring the transport properties at low energies.   
\end{abstract}

\maketitle

\section{Introduction}

Since the first evidence of  the conductance quantization in 1988 \cite{vanwees_prl_1988}, quantum point contacts (QPCs) have played a relevant role in mesoscopic physics. In fact, they naturally offer evidence of the quantum-mechanical nature of the charge carriers through a constriction whose width is comparable with the Fermi wavelength. The observation of well-defined plateaus in the conductance of the device, quantized in integer values of $2e^2/h$ (where $e$ is the electron charge and $h$ is the Planck constant), indicates ballistic transport involving a limited number of conduction channels which are spin degenerate \cite{buttiker1990}.
The quantization of $G$ can be explained from the formation of one-dimensional (1D) sub-bands in the constriction due to the lateral confinement.
Then $G$ is given by the Landauer-type formula \cite{landauer} $G=2Ne^2/h$, with $N$ the number of occupied 1D sub-bands.
A detailed analysis has shown that a variation of the gate voltage $V_G$ changes the width as well as the electron density of the constriction.
Both mechanisms move the Fermi energy $E_F$ in the channel through the 1D sub-bands and,  whenever it passes a sub-band bottom, $G$ changes by the quantized amount of $2e^2/h$. 

Ballistic constrictions are routinely fabricated in semiconducting heterostructures such as AlGaAs/GaAs ones by means of metallic split-gate which, through the application of a negative gate voltage $V_G$, forms the constriction by electrostatic depletion. Moreover, the very low carrier density ($\sim 10^{11}$ cm$^{-2}$) results in large values of the Fermi wavelength $\lambda_F$ ($\sim$ 50 nm) and the extreme cleanliness of these heterostructures also results in a large mean free path that can exceed several micrometers at low temperatures \cite{rossler}.
However, beyond these conventional materials, technological efforts have been put forward for the fabrication of nanodevices in thin films and heterostructures based on oxides exploiting the multifunctionality  and the extreme sensitivity of these materials to external perturbations \cite{Ogale}. Recently, the two-dimensional electron gas (2DEG) formed at the (001) interface  between band insulators $LaAlO_3$ and $SrTiO_3$ (LAO/STO) \cite{Ohtomo} has been much studied not only for its spin-orbit, multi-band and superconducting properties \cite{Caviglia,Caviglia2010,CavigliaReview,LevyReview}, but also for the possibility to realize nanostructures in the normal and superconducting state \cite{CavigliaQPC,LevyScience,bergeal2020}. Moreover, recent theoretical works have suggested the 2DEGs at LAO/STO interface as possible candidates for the realization of topological superconducting phases in quasi 1D nanowires \cite{Fidkowski1,Perroni,Settino}. However, in comparison with semiconductor based heterostructures, this 2-DEG  typically involves a higher carrier density ($\sim 10^{13}$ cm$^{-2}$) and has a reduced $\lambda_F$ ($\sim$ 10-50 nm) imposing stronger constraints on the practical realization of such devices. 

At the LAO/STO interface, the conduction band is formed by coupling the t$_{2g}$ 3d orbitals (d$_{xy}$, d$_{xz}$, and d$_{yz}$) at neighbouring Ti lattice sites through the 2p orbitals of the oxygen atoms. Under strong quantum confinement in the direction perpendicular to the interface, the degeneracy of the t$_{2g}$ bands is lifted, resulting in a rich and complex band structure with discrete 2D states separated by typical energies of tens of meV \cite{salluzzo2009,berner2013}. The quantization of conductance in a ballistic QPC formed by electrostatic confinement of the LAO/STO 2-DEG with a split-gate was recently demonstrated in the normal state \cite{bergeal2020}. Moreover, the measurement of the $g$-factor under a magnetic field applied in the direction normal to the interface was performed on the $n =2$ conductance step providing a value in the range 0.85-0.9, which differs from that of free electrons and is usually ascribed to the effect of strong spin-orbit coupling \cite{Caviglia2010,dagan2010}. By adding a finite source-drain bias voltage $V_B$, a comprehensive spectroscopic study of the low energy energy levels inside the QPC was performed in the depleted regime  \cite{bergeal2020}. However, while in the depleted regime only the lowest d$_{xy}$ energy bands are filled, at higher gate voltages or high carrier doping a multi-band transport can occur and stronger features related to multi-orbital nature of the subbands can emerge in the ballistic transport of an oxide based constriction.  This scenario encompasses the comprehension of two effects in the ballistic transport of oxide based QPCs: the role of the atomic spin-orbit coupling and the inversion symmetry breaking term, that can break the picture of independent orbital channels, and the multi-band transport that takes place at higher doping densities.

\begin{figure}[t!]
\centering
\includegraphics[width=\linewidth]{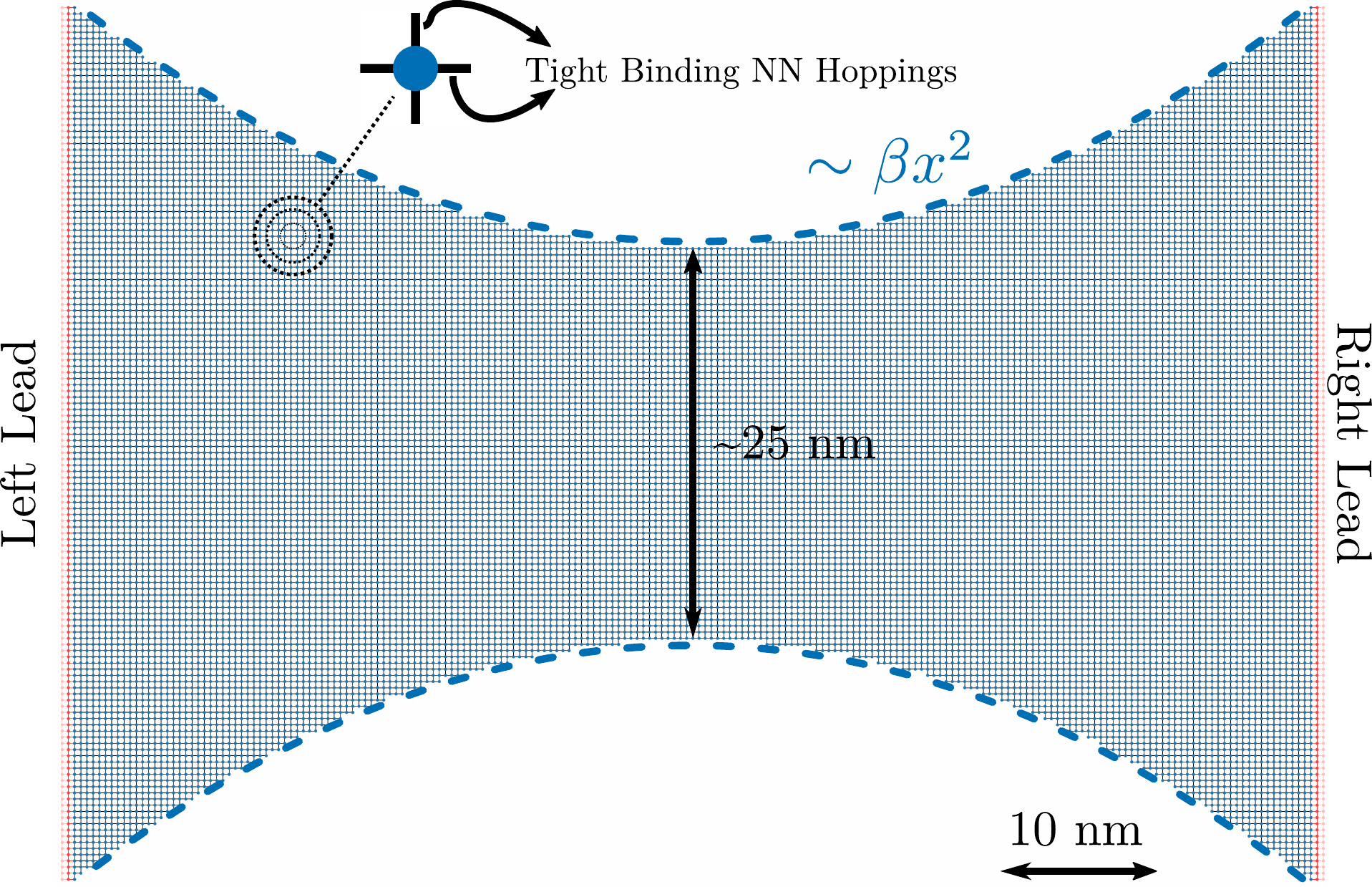}
\protect\caption{
\label{fig:Scheme} Schematic view of the QPC simulated within the TB method. The thinnest size is made of $64$ sites with a lattice constant $a=0.39 ~ nm$; the transverse parabolic hard-wall confinement has a coefficient $\beta=0.01 ~{nm}^{-1}$; the length of the TB system ensures the convergence of the results for the conductance.}
\label{Fig:1}
\end{figure}

In this paper, we theoretically analyze the ballistic transport through QPCs of multi-orbital 2DEG at the LAO/STO (001) interface modelling the normal state constriction of $25$ nm minimal size investigated in a recent experimental work \cite{bergeal2020}. In order to analyze the electronic structure of the QPC, we adopt both the $\mathbf{k\cdot p}$ approach and the tight-binding (TB) method including the effects of an applied magnetic field normal to the interface. In Fig. \ref{Fig:1}, we report a schematic view of the QPC simulated within the TB method with a curvature extracted from the recent experiment in \cite{bergeal2020}. We remark that the curvature is smooth, therefore the electron transport through the QPC is well  within the so called adiabatic regime \cite{vanwees_prl_1988,vanwees_prb_1991}. Magnetoconductance and non-linear  transport properties have been calculated in the ballistic regime with analytical approaches appropriate for an adiabatic QPC at low carrier densities and with TB based state-of-the-art numerical methods at all the densities including multi-orbital effects. 

We make a comparative analysis of the conductance, the differential conductance, and the transconductance in the absence and presence of a magnetic field normal to the interface finding three distinct ballistic transport regimes. The first one characterizes the low energy range since it involves only the  first $d_{xy}$-like sub-bands. In the absence of magnetic field, the conductance is quantized in steps of $2e^2 /h$, while, as expected, in the presence of the field, of $e^2/h$. The second transport regime entails $d_{xy}$-like sub-bands at higher energies where standard strengths of the magnetic field do not induce a splitting of the sub-bands hampering the conductance quantization in steps of $e^2 / h$. Within these two transport regimes, localized defects holes on the external sides of the QPC are able to induce conductance oscillations whose amplitude gets enhanced with increasing energy. Therefore, the conductance quantization is quite robust in the first transport regime, while it gets weakened in the second one. We point out that the theoretical results provide an accurate description for many features of the experimental data shown in \cite{bergeal2020}, including also the presence of conductance oscillations. Finally, the third transport regime is at high energies where light $d_{xy}$-like  strongly hybrydize with heavy $d_{yz}$-like, $d_{zx}$-like sub-bands so that the conductance quantization is weak against external perturbations.

The paper is organized as follows. In Section II, the model hamiltonian for an oxides strip is presented and its energy spectrum is discussed. In Section III, the low energy transport is investigated by using analytical approaches. In Section IV, the transport is examined in all the energy ranges by the TB method, with the last paragraph devoted to the analysis of the transport properties when localized defects are present on opposite sides of the QPC.   Finally, conclusions and further discussions are given in Section V.
Appendix A provides some details on the $\mathbf{k\cdot p}$ method and the analytical approach used for calculations of the transport in the limit of low energy.

\section{Model}
\label{sec:model}

The model Hamiltonian we employ for transition metal (TM) oxides describes the coupling of the t$_{2g}$ orbitals (d$_{xy}$, d$_{xz}$ and d$_{yz}$) at neighbouring Ti lattice sites through the 2p orbitals of the oxygen atoms. This TB Hamiltonian turns out to be very accurate for the description of 2D bulk electronic states \cite{CavigliaReview} containing both an atomic spin-orbit coupling and an inversion symmetry breaking term: \cite{Perroni,Zhong2013,Khalsa2013,Vivek2017,Fukaya2019}
\begin{equation}
\mathcal{H}=\sum_{\bm{k}}\Hat{D}(\bm{k})^{\dagger}H(\bm{k})\Hat{D}(\bm{k}),
\label{Htot}
\end{equation}
with
\begin{equation}
H(\bm{k})=H^0+H^\mathrm{SO}+H^{Z}+H^{M},
\end{equation}
where $\Hat{D}^{\dagger}(\bm{k})=\left[ c^{\dagger}_{yz\uparrow \bm{k}}, c^{\dagger}_{zx\uparrow \bm{k}}, c^{\dagger}_{xy\uparrow \bm{k}}, c^{\dagger}_{yz\downarrow \bm{k}}, c^{\dagger}_{zx\downarrow \bm{k}}, c^{\dagger}_{xy\downarrow \bm{k}} \right]$ is a vector whose components are associated with the electron creation operators for a given spin $\sigma$ ($\sigma=[\uparrow,\downarrow]$), orbital $\alpha$ ($\alpha=[xy,yz,zx]$), and momentum $\bm{k}$ in the 2D square Brillouin zone. Then $H^0, H^\mathrm{SO}, H^{Z}$ and $H^{M}$ represent the kinetic energy, the spin-orbit, the inversion symmetry breaking and the Zeeman interaction term, respectively. \\
In the spin-orbital basis, $H_0(\bm{k})$ is given by
\begin{align}
&H^0=\Hat{\varepsilon}_{\bm{k}} \otimes \Hat{\sigma}_{0}, \\
&\Hat{\varepsilon}_{\bm{k}}=
\begin{pmatrix}
\varepsilon_{yz} &0 &0 \\
0 & \varepsilon_{zx} &0 \\
0 &0& \varepsilon_{xy}
\end{pmatrix}, \notag \\
&\varepsilon_{yz}=2t_{1}\left( 1-\cos{k_y a }\right)+2t_{2}\left(1-\cos{k_xa}\right), \notag \\
&\varepsilon_{zx}=2t_{1}\left(1-\cos{k_x a}\right)+2t_{2}\left(1-\cos{k_y a}\right), \notag \\
&\varepsilon_{xy}=4t_{1}-2t_{1}\cos{k_x a}-2t_{1}\cos{k_y a}+\Delta_{t}, \notag
\end{align}%
where $\Hat{\sigma}_{0}$ is the unit matrix in spin space and
 $t_{1}$ and $t_{2}$ are the orbital dependent hopping amplitudes.
$\Delta_{t}$ denotes the crystal field potential as due to the symmetry lowering from cubic to tetragonal also related to inequivalent in-plane and out-of-plane TM-oxygen bond lengths.
The symmetry reduction yields a level splitting between $d_{xy}$-orbital and $d_{yz}/d_{zx}$-orbitals.

$H^\mathrm{SO}$ denotes the atomic $\bm{L} \cdot \bm{S}$ spin-orbit coupling,
\begin{align}
H^\mathrm{SO}
=\Delta_{\mathrm{SO}}\left[ \Hat{l}_x \otimes \Hat{\sigma}_x+\Hat{l}_y \otimes \Hat{\sigma}_y+\Hat{l}_z \otimes \Hat{\sigma}_z \right],
\end{align}%
with $\Hat{\sigma}_{i}(i=x,y,z)$ being the Pauli matrix in spin space and $\Hat{l}_\alpha$ ($\alpha=x,yz$) are the projections of the $L=2$ angular momentum operator onto the $t_{2g}$ subspace, i.e.,
\begin{align}
\Hat{l}_{x}&=
\begin{pmatrix}
0 & 0 & 0 \\
0 & 0 & i \\
0 & -i & 0
\end{pmatrix}, \\
\Hat{l}_{y}&=
\begin{pmatrix}
0 & 0 & -i \\
0 & 0 & 0 \\
i & 0 & 0
\end{pmatrix}, \\
\Hat{l}_{z}&=
\begin{pmatrix}
0 & i & 0 \\
-i & 0 & 0 \\
0 & 0 & 0
\end{pmatrix},
\end{align}%
assuming $\{d_{yz}, d_{zx}, d_{xy}\}$ as orbital basis.

As mentioned above, the breaking of the mirror plane, due to the out-of-plane offset of the positions of the TM and oxygen atoms, results into an inversion asymmetric orbital Rashba coupling that is described by the term $H^{Z}(\bm{k})$:
\begin{equation}
H^{Z}=\gamma \left[ \Hat{l}_y \otimes \Hat{\sigma}_0 \sin{k_x a}-\Hat{l}_x \otimes \Hat{\sigma}_{0} \sin{k_y a} \right].
\label{Hinv}
\end{equation}
This contribution provides an inter-orbital process, due to the broken inversion symmetry, that mixes $d_{xy}$ and $d_{yz}$ or $d_{zx}$.

Finally, we consider the effects of a magnetic field perpendicular to the plane of the 2DEG. The resulting Zeeman-type interaction is described by the Hamiltonian $H^{M}$, which characterizes the coupling of the electron spin and
orbital moments to the magnetic field \cite{Ruhman2014}:
\begin{equation}
H^\mathrm{M}
= M_z \left[  \Hat{l}_z \otimes \Hat{\sigma}_0 + \Hat{l}_0 \otimes \Hat{\sigma}_z \right],
\end{equation}%
\noindent with $\Hat{l}_0$ being the unit matrix in the orbital space. We notice that the inclusion of the orbital coupling to the field is only a correction which can be neglected because the spin-orbit coupling is typically larger than the strength of the applied magnetic field considered in the experiment and the Zeeman coupling to the orbital degree of freedom is less relevant.

The electronic structure at the LAO/STO (001) interface has been studied in the literature also within the $\mathbf{k\cdot p}$ approach \cite{Fasolino1,Fasolino2}, which is accurate in the limit of large wave-lengths. In this paper, we apply the $\mathbf{k\cdot p}$ procedure directly to the TB Hamiltonian in Eq. (\ref{Htot}). Actually, as discussed in Appendix A, the 2D energy eigenvalues and eigenvectors are exactly determined for small values of the wave-vector $\mathbf{k}$. In particular, in the limit of low energies \cite{Zhong2013}, due to the crystal field splitting, one can derive an effective spin Rashba interaction $\alpha$ for the $xy$-like band. Therefore, for small values of the wave-vector $\mathbf{k}$, the $d_{xy}$-like band is provided by the solutions of the following effective Rashba Hamiltonian $H_R$:
\begin{equation}
H_{R}=E_{-}+\frac{\hbar^2|\mathbf{k}|^2}{2m_1}+ \alpha (\sigma_x k_y - \sigma_y k_x)+M_z \sigma_z,
\label{HRashba}
\end{equation}
where the energy $E_{-}$ is
\begin{equation}
E_{-}=\frac{\Delta_{SO}}{2}\left( 1-\epsilon_R - \sqrt{\epsilon_R^2+2\epsilon_R+9} \right),    
\end{equation}
with $\epsilon_R=|\Delta_t|/\Delta_{SO}$, $m_1$ is the lighter mass (smaller than the free electron mass $m_e$), associated with the higher hopping $t_1$,
\begin{equation}
m_1=\frac{\hbar^2}{2t_1 a^2} \simeq 0.7 m_e,
\label{massa1}
\end{equation}
and the Rashba coupling constant $\alpha$ is determined as
\begin{equation} 
\alpha =  
\frac{ \sqrt{2} a \gamma \exp(\arcsinh \eta_R)}{[1+\exp(2 \arcsinh \eta_R)]} ,
\label{alfar}
\end{equation}
with 
\begin{equation}
\eta_{R}=\frac{(\epsilon_R+1)}{2 \sqrt{2}}.    
\end{equation}
We remark that, in the literature \cite{Perroni,Zhong2013,Ruhman2014}, the typical estimate of the Rashba coupling  corresponds to the limit of Eq. (\ref{alfar}) for very large $\epsilon_R$:
\begin{equation} 
\alpha \simeq
\frac{ \sqrt{2} a \gamma}{(2 \eta_R)} \simeq  \frac{ 2 a \gamma \Delta_{SO}}{ |\Delta_t|},
\label{alfarnew}
\end{equation}
which is clearly an overestimation. In any case, the Rashba coupling constant for the xy band depends on the atomic spin-orbit energy $\Delta_{SO}$ and the inversion symmetry breaking energy $\gamma$. 

In our work we follow the parameter setting of Ref. \cite{Perroni}: the main hopping $t_1= 300 meV$, the weaker hopping term $t_2 = 20 meV$, the atomic spin-orbit coupling $\Delta_{SO}=10 meV $, the orbital Rashba interaction $\gamma=40 meV$, and the tetragonal crystal field potential $\Delta_t=-50 meV$. Hence, one gets that an estimate of $\alpha$ in Eq.(\ref{alfar}) is given by $\alpha \simeq 4.7$ $meV \cdot nm$, a value compatible with experimental measurements \cite{CavigliaReview} in the range of typical 2DEG charge densities. In Appendix A, we show the perfect agreement between the spectrum obtained by the  Hamiltonian of Eq. (\ref{Htot}) and that from Eq. (\ref{HRashba}) for small values of the wave-vector $\mathbf{k}$. Moreover, we analyze also the behavior of the spectrum with the inclusion of the effects of the Zeeman energy $M_z$ pointing out that the description provided by the  $\mathbf{k\cdot p}$ procedure is accurate even for large magnetic fields.

\begin{figure}[t]
\centering
\includegraphics[width=8.6cm]{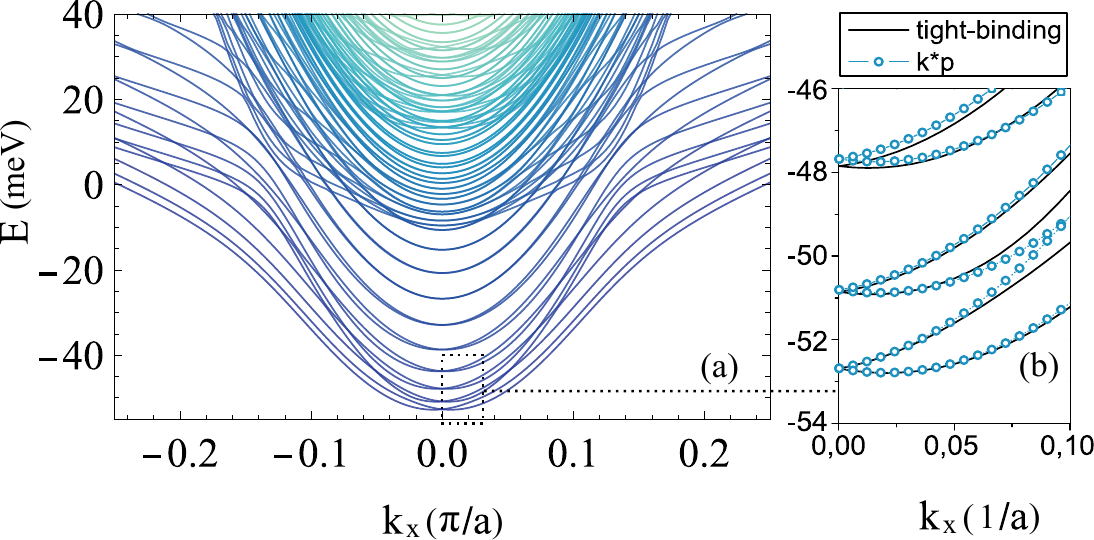}
\protect\caption{
\label{fig:Bands} Left panel: Band structure for an infinite strip with 64 transverse sites derived from the TB Hamiltonian of Eq. (\ref{Htot}). b) Right panel: $\vb k\vdot \vb p$ approximation for the lowest three sub-bands (blu circles) compared with the tight binding ones (solid black lines). }
\label{Fig:bands}
\end{figure}

Since the QPC  investigated in the experiment \cite{bergeal2020} has a minimal constriction about $25$ nm, we start analyzing an infinite strip  with $64$ transverse sites ($64 a \simeq  25$ nm, with lattice parameter $a=0.39$ nm) with hard wall boundary conditions in the transverse direction.
As shown in the left panel of Fig. \ref{Fig:bands}, the spectrum is very complex. In fact, since the  tetragonal potential $\Delta_t=-50 meV$ and the strip is not very narrow, the first sub-bands are derived from the light xy-like bands. Starting from around zero energy, the spectrum is characterized by a large superposition of sub-bands derived not only from the light xy-like bands, but also from the heavy yz-like and zx-like bands. Therefore, at high energies, the spectrum shows a quasi-continuum, while, at low energies, the xy-like sub-bands are quite distinct. Moreover, we point out that, in the left panel of 
Fig. \ref{Fig:bands}, the first three xy-like sub-bands show a clear double minimum close to ${\mathbf k}=0$, an effect that can be typically ascribed to the Rashba coupling in Rashba nanowires \cite{Perroni1,Paolino,Paolino1,Marigliano1,Marigliano2,Marigliano3}.  
However, for the next xy-like sub-bands, the double-mininum gets reduced with increasing energy. Moreover, the separation between the minima of first sub-bands follows the behavior expected for nanowires with hard wall conditions, while the next xy-like bands show even a reduction in the energy difference between the minima. These effects are obviously enhanced close to the onset of the quasi-continuum due to the contribution of heavy yz- and zx-like bands.     
Actually, the multi-orbital character of the problem is able to provide 
a complex behavior even in the case of the infinite strip.

We notice that the sub-band occupation in the experiment \cite{bergeal2020} corresponds only to the first three xy-like ones. Actually, the range of the charge densities considered in this experiment is below $10^{13}$ $cm^{-2}$. 
Therefore, as discussed in Appendix A, we adopt a $\mathbf{k\cdot p}$ procedure to analyze the properties of the first xy-like sub-bands in the strip setting up the Hamiltonian $H_{S}$ derived from Eq.(\ref{HRashba}):
\begin{equation}
H_{S}=E_{-}+\frac{\left( \hbar^2 k_x^2 +p_y^2 \right)}{2m_1}+ \alpha (\sigma_x \frac{p_y}{\hbar}  - \sigma_y k_x)+M_z \sigma_z +V(y),
\label{Hstrip}
\end{equation}
where $p_y=-i \hbar d/dy$ is the momentum operator along the y direction, and $V(y)$ imposes hard wall conditions on the transverse boundary along y (lateral size equal to $64 a$). We explicitly fix the parameters $E_{-}$, $m_1$, and $\alpha$ as derived from the 2D bulk. As discussed in Appendix A, the exact solution of $H_{S}$ can be obtained analytically for small values of $\mathbf{k}$. Particularly interesting are the effects due to the spin-orbit Rashba coupling which can be quantified through the wave-vector $k_{R}=m_1 \alpha/\hbar^2$. In fact, the single-particle spectrum of the xy-like sub-bands not only shows minima centered at $\pm k_R$, but it is globally reduced of the energy $\hbar^2 k_R^2/2 m_1$. In the next section, we will discuss how these effects determine the behavior of the conductance as a function of the gate voltage $V_G$ and the bias voltage $V_B$. As shown in the right panel of Fig. \ref{Fig:bands}, the results of the   $\vb k\vdot \vb p$ approximation are able to provide an accurate description for the lowest three xy-like sub-bands for small values of $k_x$. Tiny deviations from the TB scheme are present only for the third sub-band. We stress that $\vb k\vdot \vb p$ approximation captures the correct position of the minima of  the sub-bands, therefore, as shown in the next section, it will provide a simplified but accurate model for the transport through the adiabatic QPC. 

\section{Low energy transport}

In this section, we discuss the low-energy transport in order to analyze and clarify some aspects of recent experimental data given in Ref.\cite{bergeal2020}.   
As discussed in previous sections, within this experimental set-up, only the first three xy-like sub-bands are occupied by charge carriers which cross the QPC with $25$ nm minimal constriction and a smooth curvature. Therefore, in this section, we adopt the adiabatic approximation, briefly  discussed in Appendix A, for the treatment of the transport \cite{vanwees_prl_1988,buttiker1990,Halperin}. We remark that all the results discussed in this section have been confirmed by analogous simulations done with the more realistic TB scheme which will be exposed in the next section. 
Indeed, the scheme used in this section provides transport properties in the energy range of the first two sub-bands which are in very good agreement with the TB method.

As discussed in Appendix A, for the description of the QPC, we adopt the Hamiltonian $H_{Q}$ derived from Eq.(\ref{HRashba}):
\begin{equation}
H_{Q}=E_{-}+\frac{\left( {p_x}^2 +p_y^2 \right)}{2m_1}+ \frac{\alpha}{\hbar} (\sigma_x p_y  - \sigma_y p_x)+M_z \sigma_z + V(x,y),
\label{HQPC}
\end{equation}
where $p_x=-i \hbar d/dx$  and $p_y=-i \hbar d/dy$ are the momentum operators along the x and y direction, respectively, and the potential $V(x,y)$ imposes hard wall conditions on the x-dependent transverse boundary determined by $W(x)/2$ and $-W(x)/2$ \cite{NoteBound}. Clearly, according to the experimental set-up of Ref. \cite{bergeal2020}, $W(0)=64 a$ and $W(x)=W(0)+\beta x^2$, with $\beta \simeq 0.01$ $nm^{-1}$, that is the curvature of the QPC is small.  In this limit, the adiabatic approximation can be implemented for the treatment of Hamiltonian (\ref{HQPC}). As discussed in Appendix A, at first, the eigenvalue problem is solved in the "fast" variable y considering the "slow" variable x as fixed, then the resulting equation in the "slow" variable x is considered.  
Actually, this procedure can be analytically implemented not only in the presence of Rashba spin-orbit coupling, but also of an out-of-plane magnetic field. 

As discussed in Appendix A, we have analytically calculated the current $I$ 
as a function of the gate voltage $V_G$ and the bias voltage $V_B$ in the presence of an applied magnetic field normal to the interface.  Since many experiments show that the Rashba coupling constant $\alpha$ can change with varying the gate voltage $V_G$  \cite{CavigliaReview,LevyReview}, we consider three different values for this parameter: $\alpha=0$, $\alpha=2$ $meV \cdot nm$, which should correspond to a minimum at the LAO/STO interface, for charge densities less than $10^{13}$ $cm^{-2}$, and  $\alpha=4$ $meV\cdot nm$, which is slightly smaller than the value corresponding to typical 2DEG electron densities \cite{Caviglia2010}. We discuss transport properties both in the absence and in the presence of an applied magnetic field.

\begin{figure*}[t]
\centering
\includegraphics[width=.9\linewidth]{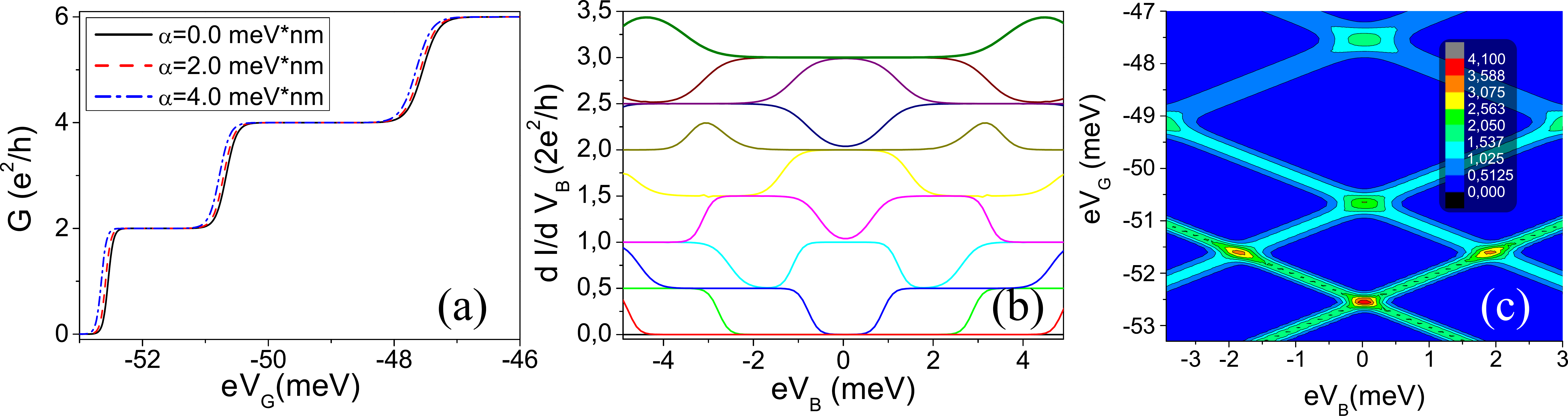}
\protect\caption{QPC in the presence of Rashba coupling. (a) Conductance $G$ of the QPC as a function of gate voltage $V_G$ for three different values of the Rashba coupling constant $\alpha$. (b) Differential conductance  as a function of bias potential $V_B$ for several values of the split-gate potential at Rashba coupling constant $\alpha=2$ $meV \cdot nm$: coloured curves correspond to values of $e V_G$ from  $-55$ meV to $-45$ meV through steps of $1$ meV. (c) Map of transconductance as a function of $V_G$ and $V_B$ at Rashba coupling constant $\alpha=2$ $meV \cdot nm$. }
\label{Fig:condRashba}
\end{figure*}

\begin{figure*}[t]
\centering
\includegraphics[width=.9\linewidth]{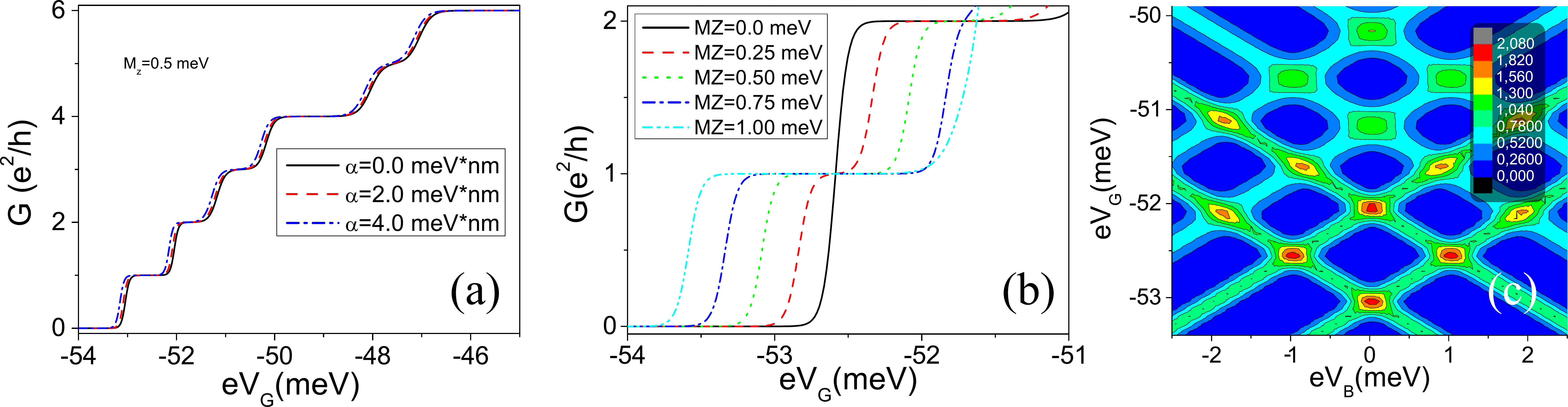}
\protect\caption{QPC in the presence of Rashba coupling and out-of-plane magnetic field. a) Conductance G as a function of gate potential $V_G$ for three different values of Rashba coupling constant $\alpha$ at magnetic energy $M_z=0.5$ meV. b) Conductance G as a function of gate potential $V_G$  for several values of the magnetic energy $M_z$ at Rashba coupling $\alpha=2.0$ $meV \cdot nm$. c) Map of transconductance as a function of $V_G$ and $V_B$ at magnetic energy $M_z=0.5$ meV at Rashba coupling $\alpha=2.0$ $meV \cdot nm$.}
\label{Fig:condRashba2}
\end{figure*}

First, we discuss the effects of the spin-orbit Rashba coupling $\alpha$ on the the behavior of the conductance $G$ (defined as $\partial I / \partial V_B$ in the limit $V_B \rightarrow 0^+$) as a function of the gate potential $V_G$. As reported in the left panel of Fig. \ref{Fig:condRashba}, the conductance shows plateaus corresponding to quantized values of the conductance quantum $G_0 = 2e^2/h$ in agreement with experimental results \cite{bergeal2020} which indicate a ballistic transport in the QPC. From the comparison with low energy bands in Fig. \ref{Fig:bands}, we notice that the onset of the plateaus does not precisely coincide with the minima of the bands. Actually, as discussed in Appendix A, the smoothing of the step between two consecutive plateaus is due to the small, but finite spatial curvature of the adiabatic QPC shown in Fig. \ref{Fig:1}. We remark that the value of the curvature is extracted from the experimental set-up in \cite{bergeal2020}, and that the Rashba coupling provides a small, but finite shift of the conductance curves in agreement with the effects discussed in Fig. \ref{Fig:bands} about the energy spectrum. Additional effects are not expected in the low energy regime since the Rashba term is proportional to the momentum operator, therefore to the first derivative of the wave function which is slowly changing in the adiabatic QPC with increasing the coordinate x. In the next section, however, we will find that effects due to the atomic spin-orbit and the inversion symmetry breaking term get enhanced with increasing the energy range.

\begin{figure*}[t]
\centering
\includegraphics[width=0.85\textwidth]{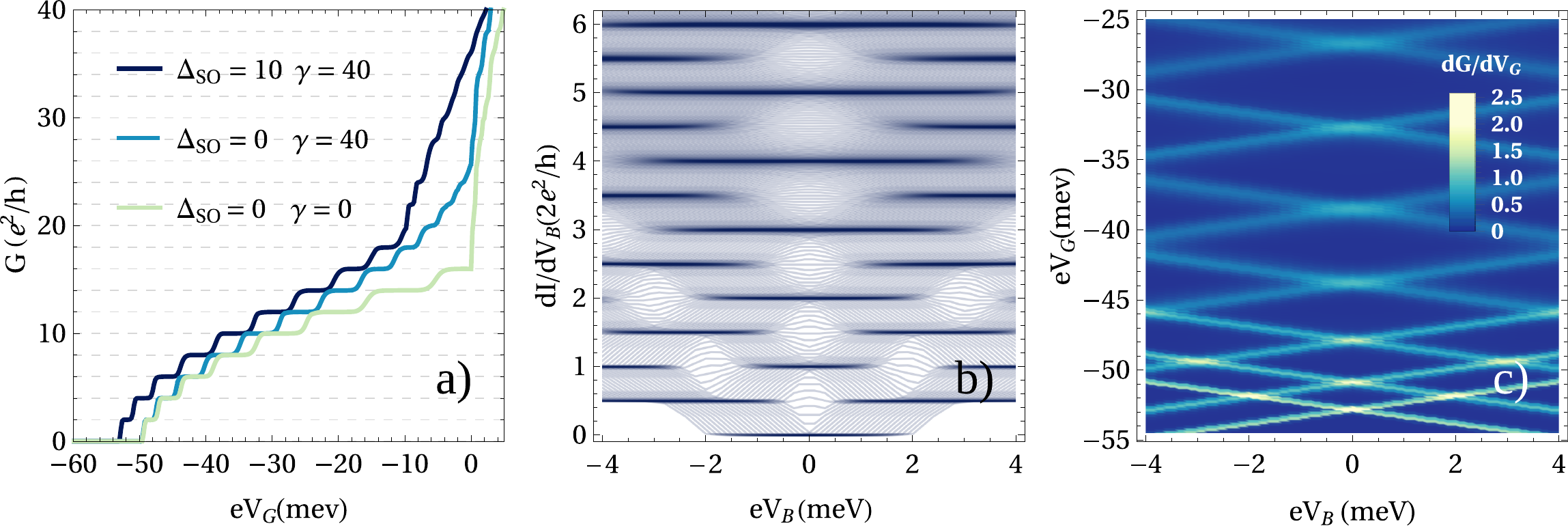}
\caption{
\label{fig:High1}
a) Conductance of the QPC as a function of energy for three different set of Hamiltonian parameters, as indicated in the legend. b) Differential conductance as a function of $V_B$ for several values of the split-gate potential. c) Map of the transconductance as a function of $V_G$ and $V_B$.} 
\label{Fig:condMultSpinorbit}
\end{figure*}

\begin{figure*}[t]
\centering
\includegraphics[width=0.85\textwidth]{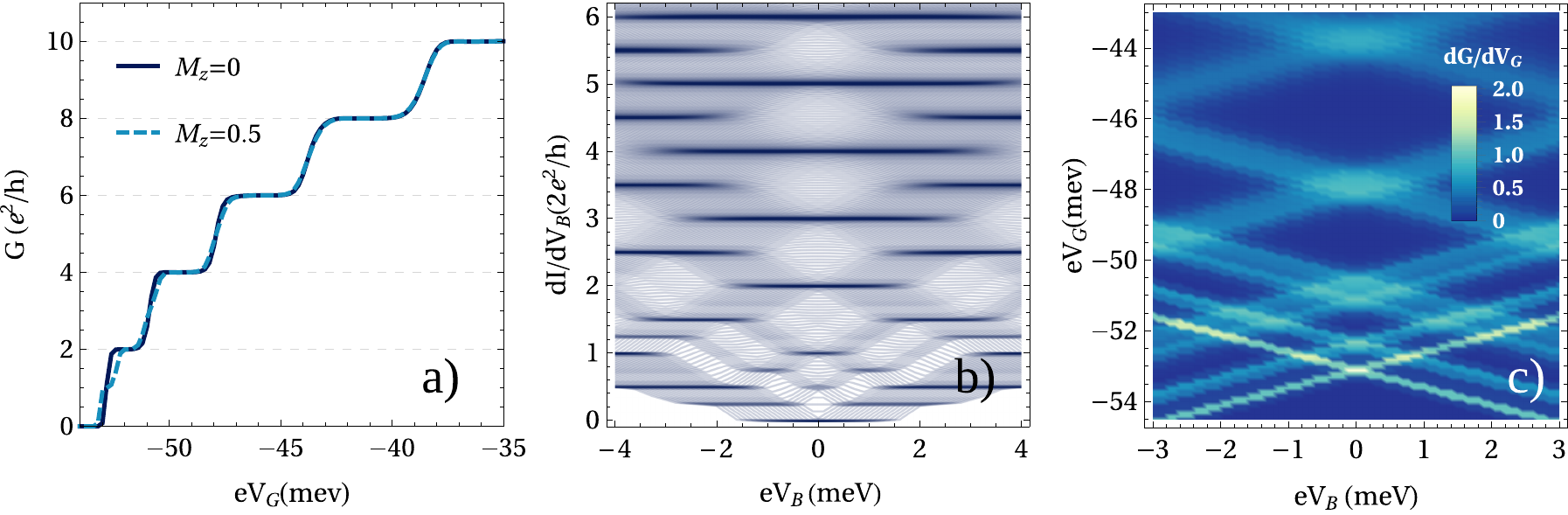}
\caption{
\label{fig:High2} a) Conductance of the QPC as a function of energy in presence of a out-of-plane magnetic field. b) Differential conductance as a function of $V_B$ for several values of the split-gate potential at the magnetic energy $M_z=0.5$ meV. c) Map of transconductance as a function of $V_G$ and $V_B$ at the magnetic energy $M_z=0.5$ meV} 
\label{Fig:condMultSpinorbit}
\end{figure*}

For the discussion of the differential conductance and the transconductance in the absence of magnetic field, we consider a Rashba coupling constant $\alpha=2$ $meV \cdot nm$, since it is close to experimental estimates in the limit of low charge density. In the middle panel of Fig. \ref{Fig:condRashba}, we plot the differential conductance (defined as the derivative of the current respect to $V_B$)  as a function of bias potential $V_B$ for several values of the split-gate potential. The differential conductance presents plateaus not only at $n G_0$ but also at  $(n+1/2) G_0$. Actually, additional plateaus at $(n+1/2) G_0$ can be obtained when the number of modes available for left-going and right-going charge carriers differs by one \cite{Kouwenhoven,Patel,Glazman}. This effect takes place when the energy 
$e V_B$ exceeds the splitting between two sub-bands. Moreover, in our 2D model, at odds with experimental results, we find a perfect  symmetry  between positive and negative values of bias potential $V_B$. 

In the left panel of Fig. \ref{Fig:condRashba}, we show contour plots of the transconductance $\partial^2 I / \partial V_G \partial V_B$  as a function of bias potential $V_B$ and gate potential $V_G$ in the absence of magnetic field. Therefore, we analyze again the non-linear transport emphasizing now the transitions between the conductance plateaus. The maxima of the transconductance form a diamond-like structure centered at  $V_B=0$. Apparently, the maxima decrease with increasing the energy, while, at the same time, the diamond-like structure gets enlarged. Therefore, the edges of the diamonds provide directly the separation between two consecutive energy levels. We point out that the results for the differential conductance and the transconductance are in good agreement with experiment \cite{bergeal2020} in the absence of magnetic field. 

We investigate now the effects of a magnetic field perpendicular to the 2-DEG plane on the transport properties. As seen in left panel of Fig. \ref{Fig:condRashba2}, the plateaus of conductance are quantized in half-integer values of $G_0$, indicating that the spin degeneracy is lifted. In analogy with Fig. \ref{Fig:condRashba2}, we first analyze the effects of three different values  of Rashba coupling, then, we consider $\alpha=2.0$ $meV \cdot nm$ for the study of further transport properties. In particular, in the middle panel of Fig. \ref{Fig:condRashba2}, we plot the conductance $G$  for several values of the magnetic energy $M_z$ focusing on the halving of the step. Apparently, the additional plateau at half $G_0$ gets enhanced with increasing the strength of the magnetic field. Finally, in the right panel of Fig. \ref{Fig:condRashba2}, we show contour plots of the transconductance as a function of $V_G$ and $V_B$ in the presence of the magnetic field. Seemingly, making the comparison with the right panel of Fig. \ref{Fig:condRashba2}, the magnetic field involves a duplication of the diamond-like structure. We remark that most features of the conductance and the transconductance are in good agreement with experimental results \cite{bergeal2020} even in the presence of a magnetic field.

\section{Multi-orbital transport}

In this section we describe the electronic transport properties of the QPC within the TB approximation by numerically implementing a finite-size system described by the Hamiltonian (\ref{Htot}). This description allows us to go beyond the low energy regime, that has been significantly characterized by the 
$ \vb k \vdot  \vb p$ method. We will show the TB approach is necessary to get insights on the multi-orbital ballistic transport characteristic of higher doping densities. The conductance of the QPC described in Fig. \ref{fig:Scheme}) is calculated within the TB method by using KWANT \cite{Kwant} and NumPy\cite{NumPy} libraries.

By looking at Fig. \ref{fig:High1} a) we see that while the atomic spin-orbit term mainly introduces an orbital-dependent energy shift, the orbital Rashba coupling, weighted by $\gamma$, acts more intriguingly: low energy subbands are left unchanged while high energy ones are squeezed toward negative energies. The width of the plateaus is almost constant or decreasing, while it should be expected to be linearly increasing because of the hard wall confinement in the transverse direction. Indeed, it is possible to see in Fig. \ref{fig:Bands} that the distance between the lowest type of sub-bands (related to the orbital $d_{xy}$) is not increasing. Therefore, hybridization at higher energy between sub-bands with similar orbital character is able to change the energy spectrum and the conductance of the adiabatic QPC. Finally, we notice that at higher charge densities, where the hybridization takes place between all three orbitals, the plateaus are very narrow. At the highest densities considered in this paper, the plateaus completely disappear, so that the conductance curve does not show any quantization. In fact, as seen in Fig. \ref{fig:Bands}, the yz- and zx-like sub-bands are heavier than the xy-like sub-bands, therefore, they gather more easily at energies of the order of $0-40$ meV creating a quasi-continuum.

In analogy with the results reported in \ref{Fig:condRashba2},  Fig. \ref{fig:High1} b) and c) show the differential conductance and the map of the transconductance. The differential conductance presents plateaus at $n G_0$ and $(n+1/2) G_0$ in a large range of energies. Furthermore, the  symmetry  between positive and negative values of bias potential $V_B$ is kept for all the sub-bands analyzed in the paper. Likewise, the transconductance shows maxima forming a diamond-like structure centered at  $V_B=0$ up to large energies. 
The diamonds are visible up to energies of the order of $-25$ meV, hence well beyond the range of energies studied in the previous section. Indeed, some features of the non-linear transport are similar to those found in the low eneergy limit. However, both pictures emphasize the non-increasing energy distance between sub-bands determined by the orbital Rashba coupling $\gamma$.

Finally, we analyze the effects of the magnetic field on the linear and non-linear transport properties in the energy range where xy-like sub-bands provide the relevant spectrum. Quite surprisingly, the effect of the turning-on of an out-of-plane magnetic field is only visible for low energies. Indeed, in complete analogy with the results discussed in the previous section, we see in Fig. \ref{fig:High2} a) that the expected halving of plateaus occurs only for the first two or three sub-bands. On the other hand, higher energy  plateaus are only smoothed, as visible also from the differential conductance in Fig. \ref{fig:High2} b) and the transconductance in \ref{fig:High2} c). We stress that this effect is due to the strong interplay between the orbital Rashba coupling $\gamma$, the atomic spin-orbit energy $\Delta_{SO}$ and the Zeeman term $M_Z$. In fact, with increasing the energy, the splittings induced by the magnetic field are quenched, therefore the sub-band spectrum is strongly affected by other energy terms mixing different sub-bands and orbitals.

\paragraph*{Effects of localized defects on the conductance.}

\begin{figure}[h]
\centering
\includegraphics[width=0.9\linewidth]{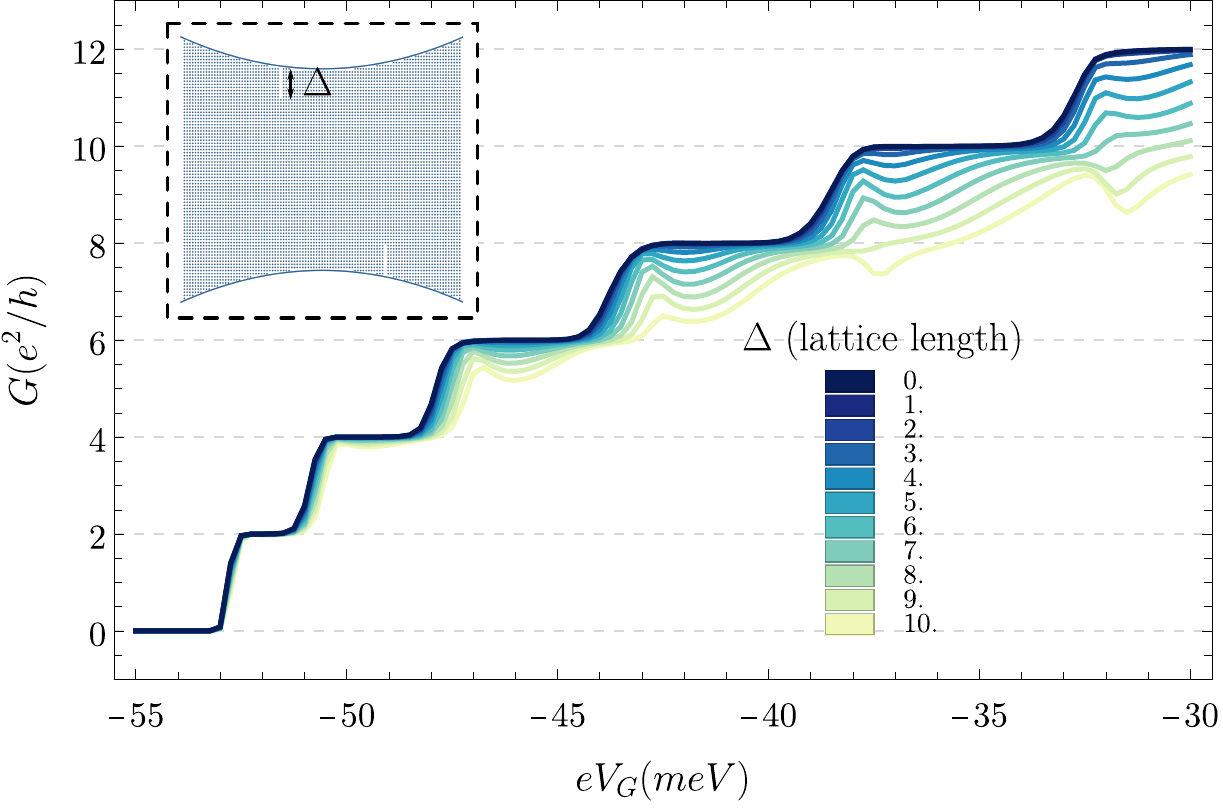}
\protect\caption{Conductance G of the QPC as a function of the gate potential $V_G$ for different lengths $\Delta$ of the holes on the two sides of the QPC.  In the inset the position of the two defects, one on top side of the QPC, the other on the down side.}
\label{Fig:A3}
\end{figure}

In this paragraph, we  analyze effects associated with the geometry of the constriction, in particular, we focus on the presence of localized defects present on  opposite sides of the QPC (see Fig. \ref{Fig:A3}). This analysis is motivated by the fact the oscillations of the conductance can be experimentally observed on the plateaus \cite{bergeal2020}. Indeed, we show that defects are able to induce conductance oscillations whose amplitude gets enhanced with increasing energy.

First, we have considered the effect of an extended spatial modulation on the border of the QPC. We have found that this perturbation is not able to strongly affect the conductance. Next, we have examined the effects of two localized defects close to opposite borders of the QPC as shown in the inset of Fig. \ref{Fig:A3}: two pin-like irregularities with variable depth $\Delta$. In fact, the couple of defects induces a progressive reduction of the value of the conductance at the plateaus with increasing the energy. Moreover, as seen in Fig. \ref{Fig:A3}, this value of the conductance gets lowered with increasing the length of the defect. 
Finally, this reduction is accompanied also by numerous oscillations of the conductance. We have checked that neither the atomic spin orbit coupling nor the asymmetric Rashba coupling are able to hamper the decreasing of the value of the conductance and its oscillating behavior, which 
can be entirely ascribed to interference effects between the two holes. This paragraph completes our work which hence provides a very accurate description of the ballistic transport in all the regimes.

\section{Conclusions and Discussions}
In this paper, we have analyzed linear and non-linear ballistic transport properties  within the Landauer-B\"uttiker method  using the tight-binding model and the $\mathbf{k\cdot p}$ approach for the electronic structure of an adiabatic QPC at the (001) LAO/STO interface. We have focused on the interplay between the atomic spin-orbit coupling, the inversion symmetry breaking term and the out-of-plane magnetic field pointing out three transport regimes. The first one takes into account the  first xy-like sub-bands, therefore it is limited to low energies. In this regime, the quantization of the conductance is quite marked both in the absence and presence of the magnetic field. The second transport regime is found at intermediate energies, since it involves further $d_{xy}$-like sub-bands. In this regime, the conductance quantization is less pronounced  since typical strengths of the magnetic field are not able to induce the splitting of the sub-bands hindering the increase of transport channels. Moreover,  due the orbital Rashba coupling $\gamma$, the energy distance between sub-bands does not increase with increasing the energy at odds with what occurs for the first xy-subbands. Finally, the third transport regime is characterized by the mixing between light $d_{xy}$-like and heavy $d_{yz}$-like, $d_{zx}$-like sub-bands so that the conductance plateaus become very narrow indicating that the conductance quantization is weak. 

This work has been partly motivated by recent experiments \cite{bergeal2020} exploring low energy transport in the normal state. In particular, in experimental data,  oscillations of the conductance can be observed on the plateaus. In order to address this issue, we have analyzed the effects of localized  pin-like defects present on external sides of the QPC. Indeed, it is found that localized defects are able to induce conductance oscillations whose amplitude gets enhanced with increasing energy. As a result of our analysis, these oscillations can be ascribed to interference effects associated with the geometry of the constriction. 


In this paper, the focus has been on the transport properties of the normal state. Therefore, this work represents  only the starting point for the theoretical analysis of superconducting QPCs which can be very relevant for addressing the possibility of topological superconductivity and Majorana fermions \cite{Wimmer} at the  LAO/STO interface \cite{Perroni,Settino,Maiellaro}. Work in this direction is in progress.


\begin{acknowledgements}
We acknowledge useful discussions with Mario Cuoco, Fiona Forte and Marco Salluzzo.
This work was supported by the project Quantox Grant Agreement No. 731473,  QuantERA-NET Cofund in Quantum Technologies, implemented within the EU-H2020 Programme.
\end{acknowledgements}

\clearpage
\begin{appendix}

\section{Methods for continuum models}
In this Appendix we briefly discuss the $\vb{k} \vdot \vb{p}$ approach used for analytical calculations of the electronic structure of the 2D bulk, the strip and the QPC.  Finally, we provide some details on the calculation of the conductance of the adiabatic QPC starting from the 
$\vb{k} \vdot \vb{p}$  energy levels.

\subsection{$\vb{k} \vdot \vb{p}$ method for  2D bulk}

\begin{figure}[H]
\centering
\includegraphics[width=1.1\linewidth]{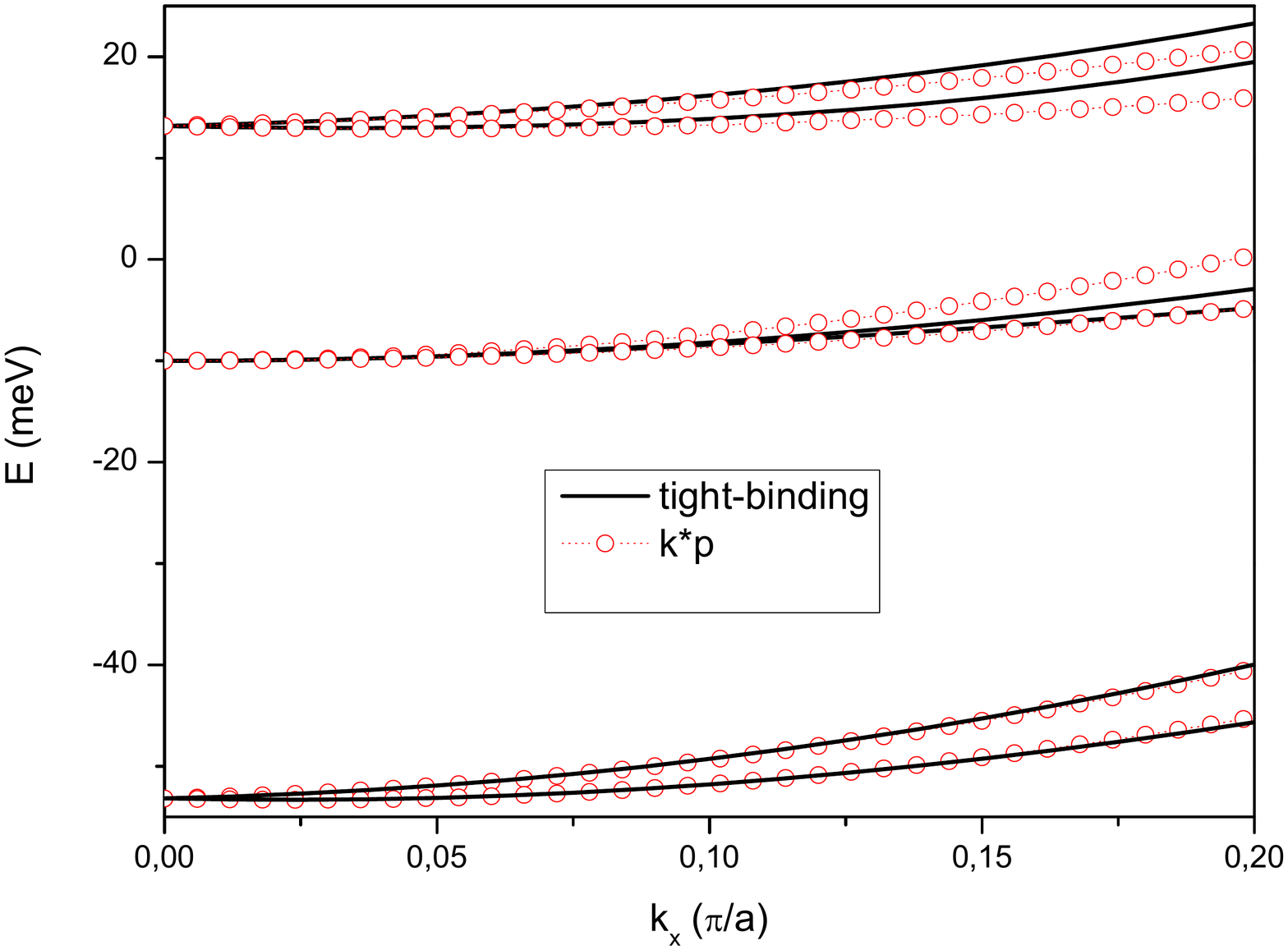}
\protect\caption{2D energy eigenvalues (in units of meV)  as a function of the  wave vector $k_x$ (in units of $\pi/a$): comparison between the results in the $\vb{k} \vdot \vb{p}$ and the TB method.}
\label{Fig:A1}
\end{figure}

In this subsection,  we apply the $\mathbf{k\cdot p}$ procedure directly to the TB 2D Hamiltonian in Eq. (\ref{Htot}) getting  energy eigenvalues and eigenvectors which are exact for small values of the wave-vector $\mathbf{k}$. In particular, at $\mathbf{k}=0$, the inversion symmetry breaking term (\ref{Hinv}) vanishes, therefore the TB Hamiltonian can be analytically diagonalized, providing, in the case of zero magnetic field, the eigenvalues $E_{\pm}$ 
\begin{equation}
E_{\pm}=\frac{\Delta_{SO}}{2}\left( 1-\epsilon_R \pm \sqrt{\epsilon_R^2+2\epsilon_R+9} \right),    
\end{equation}
with $\epsilon_R=|\Delta_t|/\Delta_{SO}$, and the eigenvalue $E_0=-\Delta_{S0}$.
Then, one considers the matrix elements of the $\mathbf{k}$-dependent inversion symmetry breaking term and  the kinetic energy between the exact eigenstates obtained at $\mathbf{k}=0$ finding approximate eigenvalues at finite $\mathbf{k}$.  

We recall that, in this paper, the following parameters are used for the 2DEG at LAO/STO (001) interface \cite{Perroni} with typical carrier densities: $t_1=300$ meV, $t_2=20$ meV, $\Delta_{SO}=10$ meV, $\gamma=40$ meV, $\Delta_t=-50$ meV. Therefore, as shown in Fig. \ref{Fig:A1}, at $\mathbf{k}=0$, one gets three eigenvalues which are doubly degenerate. Then, the degeneracy is broken at finite $\mathbf{k}$. In particular, we point out the different behaviors as a function of $\mathbf{k}=0$ for the first xy-like and the third zx-like doublet. This behavior indicates the possibility to describe the spectrum in terms of an emergent Rashba coupling constant $\alpha$, which, for the first doublet, has been specified in previous sections of the manuscript. For the first xy-like doublet, the term quadratic in the wave vector depends on the mass $m_1$ defined in Eq. (\ref{massa1}).  
On the other hand, the second doublet corresponding to yz-like bands starts at the atomic energy $E_0$ and cannot be interpreted in terms of a Rashba term linear in the wave vector \cite{Perroni}.

\begin{figure}[H]
\centering
\includegraphics[width=1.1\linewidth]{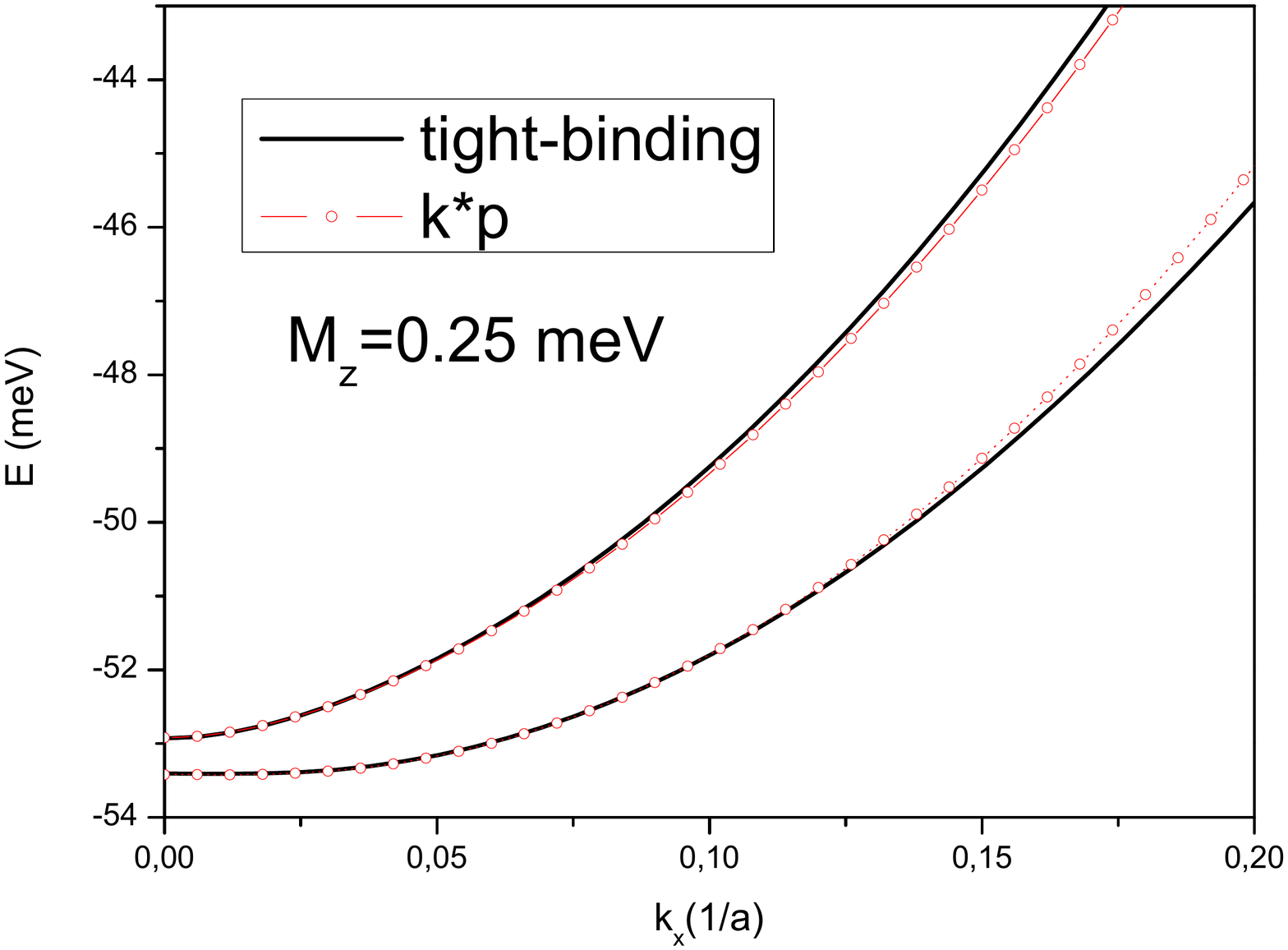}
\protect\caption{2D energy eigenvalues (in units of meV)  as a function of the  wave vector $k_x$ (in units of $\pi/a$) for magnetic energy $M_z=0.25$ meV: comparison between the results in the $\vb{k} \vdot \vb{p}$ and the TB method.}
\label{Fig:A2}
\end{figure}

Finally, we use the $\vb{k} \vdot \vb{p}$ method to get the 2D band structure in the presence of an out-of-plane magnetic field. Seemingly, the method is able to grasp the gap at $\mathbf{k}=0$  induced by the applied magnetic field and provide the bands at finite but small values of the wave vector.

\subsection{$\vb{k} \vdot \vb{p}$ method for strip}

We adopt the $\mathbf{k\cdot p}$ procedure to analyze the properties of the first xy-like sub-bands in the strip. To this aim, we explicitly consider the parameters $E_{-}$, $m_1$, and $\alpha$ as derived from $\mathbf{k\cdot p}$ procedure in the 2D bulk. Once fixed these parameters, the corresponding Hamiltonian for the strip given in Eq. (\ref{Hstrip}) can be exactly solved at $k_x=0$ and complex matrix elements of the hamiltonian operator (at small $k_x$) between the eigen-vectors at $k_x=0$ can be analytically evaluated. This procedure provides an accurate description for the lowest three xy-like sub-bands for small values of $k_x$.

\subsection{Adiabatic approximation for the QPC}

As before, we explicitly consider the parameters $E_{-}$, $m_1$, and $\alpha$ as derived from $\mathbf{k\cdot p}$ procedure in the 2D bulk. Therefore, we use the $\mathbf{k\cdot p}$ procedure to set up the Hamiltonian given in Eq. (\ref{HQPC}) which is solved exploiting the small curvature of the QPC. Actually,  the eigenvalue problem is solved in the "fast" variable y considering the "slow" variable x as fixed getting the x-dependent energy levels $E_n(x)$, with $n$ energy label. Then, the resulting equation in the "slow" variable x is considered both in the absence and presence of an out-of-plane magnetic field including the first terms of the expansion of  $E_n(x)$ in the variable x:
$E_n(x) \simeq E_{n0}-\omega_n^2 x^2/2$, with $\omega_n$ determined by the small curvature of the adiabatic QPC.

Once known the energy levels $E_{n0}$ of the QPC and the level dependent quantities $\omega_n$, using the Landauer-B\"uttiker approach \cite{buttiker1990,Halperin}, the contribution of each channel to the conductance of the QPC is given by its transmission $T_n$:
\begin{equation}
T_n(E)=\frac{1}{1+exp(-\pi \epsilon_n)},
\end{equation}
where $\epsilon_n=2 (E-E_n0)/\hbar \omega_n$. The total transmission is the sum of the transmission of each of the channels. Therefore, the quantities $\omega_n$ provide a smoothing of the transmission steps depending on the curvature of the QPC. Finally, from the transmission, we have calculated the current $I$ as a function of the gate voltage $V_G$ and the bias voltage $V_B$ in the presence of an applied magnetic field normal to the interface.

\end{appendix}


\end{document}